 
  \magnification=\magstep1

   \font\stephalf=  cmr6 scaled\magstep0  
 \font\abstract=  cmr7 scaled\magstep1

\settabs 18 \columns
 \hsize=16truecm
  
  \input epsf

\def\s{\sigma}

\def\b{\bigskip}
\def\bb{\bigskip\bigskip}

\def\no{\noindent}
\def\r{\rightline}
\def\ce{\centerline}
\def\ve{\vfill\eject}

\def\r{\rightline}

\def\s{\sigma}

\def\harr#1#2{\smash{\mathop{\hbox to .25 in{\rightarrowfill}}
 \limits^{\scriptstyle#1}_{\scriptstyle#2}}}

\def\today{\ifcase\month\or January\or February\or March\or April\or
May\or June\or July\or
August\or September\or October\or November\or  December\fi
\space\number\day, \number\year }

\r \today
\bb\bb\bb



\def\e{\eta}

\def\p{\partial}

\def\sqr#1#2{{\vcenter{\vbox{\hrule height.#2pt
\hbox{\vrule width.#2pt height#2pt \kern#2pt
\vrule width.#2pt}
\hrule height.#2pt}}}}

 \def\1/2{{\scriptstyle{1\over 2}}}
 \def\a/2{{\scriptstyle{3\over 2}}}
 \def\5/2{{\scriptstyle{5\over 2}}}
 \def\7/2{{\scriptstyle{7\over 2}}}*
 \def\3/4{{\scriptstyle{3\over 4}}}

\def\picture #1 by #2 (#3){
  \vbox to #2{
    \hrule width #1 height 0pt depth 0pt
    \vfill
    \special{picture #3} 
    }
  }

\def\scaledpicture #1 by #2 (#3 scaled #4){{
  \dimen0=#1 \dimen1=#2
  \divide\dimen0 by 1000 \multiply\dimen0 by #4
  \divide\dimen1 by 1000 \multiply\dimen1 by #4
  \picture \dimen0 by \dimen1 (#3 scaled #4)}
  }

%
%

\font\steptwo=cmb10 scaled\magstep2
\font\stepthree=cmb10 scaled\magstep4
\magnification=\magstep1

  {\ce {\stepthree Ideal Stars and General Relativity  }} 
\b
 \ce{Christian Fr\o nsdal}
\b
 \ce{\it Physics Department, University of California, Los Angeles CA
 90095-1547 USA}
\vskip.5in

\def\sqr#1#2{{\vcenter{\vbox{\hrule height.#2pt
\hbox{\vrule width.#2pt height#2pt \kern#2pt
\vrule width.#2pt}
\hrule height.#2pt}}}}

\def \r{\rightarrow}

\def\e{{\rm e}}
{\abstract 
\no{\it ABSTRACT.} We study a system of differential equations that
governs  the  distribution of matter in the theory of General Relativity. The
new element in this paper is the use of a dynamical action principle that
includes all the degrees of freedom, matter as well as metric. The matter
lagrangian defines a relativistic version of non-viscous, isentropic 
hydrodynamics.  The matter fields are a scalar density and a velocity
potential;  the conventional, four-vector velocity field is replaced by the
gradient of the potential and its scale is fixed
by one of the eulerian equations of motion, an   innovation that significantly
affects the imposition of boundary conditions.   If the density is integrable at
infinity, then the metric approaches the Schwarzschild metric at large
distances. There are stars without boundary and with finite total mass; the
metric shows rapid variation in the neighbourhood of the Schwarzschild radius
and there is a very small core where a singularity indicates that the gas laws
break down. For stars with boundary there emerges a new, critical relation
between the radius and the gravitational mass, a  consequence of the stronger
boundary conditions.   Tentative applications are suggested, to certain Red
Giants, and to neutron stars, but the investigation reported here was limited to
polytropic equations of state. Comparison with the results of Oppenheimer and
Volkoff on neutron cores shows a close agreement of numerical results. However,
in  the model the boundary of the star is fixed uniquely by the required
matching of the interior metric to the  external Schwarzschild metric, which is
not the case in the traditional approach.
  There are solutions for
which the metric is very close to the Schwarzshild metric everywhere outside the
horizon, where the source is concentrated. The Schwarzschild metric is
interpreted as the metric of an ideal, limiting configuration of matter, not as
the metric of empty space.} 

 \b\b
\no{\steptwo 1. Introduction} 
\b
\ce{\bf Background }

Relativistic hydrodynamics, as a dynamical theory,  has not yet been fully
developed. Modern textbooks on General Relativity all present the same intuitive
idea of a phenomenological fusion of Classical Thermodynamics and General
Relativity, often without encouraging a
skeptical attitude. In this paper we shall try to formulate a dynamical theory
of matter interacting with the metric, leaving the thermodynamic aspects to be
studied within the framework of the theory. As it turns out, it will be
necessary to challenge some traditional concepts, which we hope will open a
discussion and perhaps indicate a direction of a future theory of relativistic
hydrodynamics.    

Our reluctance to entirely embrace some of the
well known paradigms that supplement Einstein's General Relativity 
centers on  
Tolman's famous formula for the source term in Einstein's field equations,
$G_{\mu\nu} =  8\pi GT_{\mu\nu}$, namely
$$
T_{\mu\nu} = (\hat \rho + \hat p)\, U_\mu U_\nu - g_{\mu\nu} \hat p.
$$
(The density $\hat \rho$ and the pressure $\hat p$ must not be identified with
the scalar fields $\rho$ and $p$ introduced below.)
  To apply this formula one requires additional input
in the form of an equation of state   that relates  
$\hat p$ to  $\hat\rho$. And something must be done to pin down the 4-velocity
field
$U$. The question of how to do that is a central issue of this paper.
\b

\ce{\bf The proposal}

We propose to limit the investigation to irrotational flows, introducing
a local velocity potential,  as is done in certain applications of  
non-relativistic hydrodynamics; the vector field $U$ is replaced by $\Psi$, 
$\Psi^\mu =   g^{\mu\nu} 
\psi_{,\nu}$. There is precedence for this, in Jeans' paper [J] of 1902. 
Believing that any theory gains internal coherence and plausibility by being
formulated as a  variational problem, we introduce a dynamical action principle,
including a matter lagrangian. It is a model, too simple to be realistic, but it
is a direct generalization of non-relativistic isentropic hydrodynamics; it does
not violate any general physical principles, and there is nothing to suggest
that it should be unsuitable for describing certain configurations of an actual
physical system. It will be applied to the problem of the equilibrium of
idealized spherical stars and, later, to the question of their stability with
respect to radial oscillations. An application to cosmology is in preparation.
 
The Euler-Lagrange equations of the model include a conservation law 
$$
\partial_\mu J^\mu = 0,~~ J^\mu = \sqrt{-g\,}\rho \,\Psi^\mu.
$$
They also impose a restriction on $g_{\mu\nu}\Psi^\mu \Psi^\nu$,
alternatively derivable (in a slightly weaker form) from the (contracted) Bianchi
identities and the conservation law,  but they do not fix this quantity at unity.
In a dynamical context we expect that all
fields are governed by equations derived from the action principle and then it is
not possible to impose normalization conditions in an {\it ad hoc}
manner.  Nor is it advisable to define the action principle in terms of a
restricted class of variations that respect an {\it a priori} normalization,
as was advocated  in the papers   [SW], [S] and [Ta].
  It is of course possible
to define a normalized vector field $U$ by rescaling of the gradient field
$\Psi$ and adopting Tolman's formula for $T_{\mu\nu}$ in terms of $U,\hat\rho$ 
and
$\hat p$. The question is whether these are the fields that are
related by an equation of state.

\ve
\ce{\bf Outline and conclusions }

 The model, introduced in Section 2,
is based on the following matter contribution to the action  of General
Relativity,
$$
A = \int d^4x\sqrt{-g}\Big({\rho\over 2}( g^{\mu\nu} \psi_{,\mu}
\psi_{,\nu} - c^2) - V[\rho]\Big) =:  \int d^4x\sqrt{-g}~{\cal
L}~.\eqno(1.1)
$$
It involves 2 scalar fields, a ``density"  $\rho$ and a velocity potential
$\psi$. The associated energy momentum tensor is used as source for
Einstein's equations.   The metric is limited
throughout this paper to a class that is characterized by rotational symmetry
and the existence of coordinates
$t,r,\theta,\phi$ such that the line element takes the form  
 $$
(ds)^2 = {\e}^\nu(dt)^2 -  \e^\lambda(dr)^2  - r^2d\Omega^2, \eqno(1.2)
$$
with $g_{tt} = \e^\nu$ and $g_{rr} = \e^\lambda$ depending only on $r$
and $t$. A remark  at the end of the section
points  out differences between the conventional approach and the
point of view to which we are led by the model. 

Section 3 deals with the static limit or equilibrium configuration, beginning
with a radically new interpretation of the Schwarzschild metric. It
is not the metric of everywhere empty space (which is Minkowski) but
instead the metric of a limiting configuration of matter, concentrated at
the horizon. This interpretation of the Schwarzschild metric is supported by
another argument. Let us approach the  Schwarzschild problem from the point of
view of newtonian gravity. We know, for we teach this to our undergraduates,
that a spherical shell of matter (at $r = R$) gives rise to no gravitational
field on the inside. Indeed the potential is
$$
\phi(r) = -{ MG\over r}\theta(r-R) -{MG\over R}\theta(R-r).\eqno(1.3)
$$
We get this result by viewing the determination of the potential on the
outside, respectively the inside, as two separate problems.  It would 
  not occur to us propose that the function 
$-MG/r$ that represents the potential on the outside,   
analytically continued to the region $r<R$, has anything to do with the
problem, unless, of course, we were ignorant of the distribution of mass.
Unfortunately, that is precisely what was done after Schwarzschild presented his
solution for the outside.

Also in this section (Section 3) we introduce our choice of the functional
$V[\rho]$ that will be used for all our calculations,  
  the simple form $V[\rho] = a\rho^\gamma$, $a$ and $\gamma$
constant. There are strong internal indications that
$V$ may be interpreted as a relativistic analogue of the internal energy per unit
mass, then the Lagrangian density ${\cal L}$ is the pressure (see [S] and
[Ta]) and we obtain the isentropic or polytropic equation of state
$$
  p = a(\gamma-1) \rho^\gamma.
$$
The number $\gamma$ would thus be interpreted as the ratio
$C_p/C_v$ of heat capacities at constant pressure and volume.
This number has been taken to be constant, or nearly so, or at
least piecewise constant, in all works on stellar dynamics that we are acquainted
with.  
The choice of density to be entered into the equation of state varies
(particle number density, energy density, density of free electrons ...). Thus
Oppenheimer and Volkoff take over Chandrasekhar's equation
of state [C] but substitute energy density for mass density [OV].  

Our point of view is that taking $V[\rho] = a\rho^\gamma$ is an attractive 
choice for $V$, successful in the non-relativistic limit, and suitable for a
preliminary exploration of the physical systems within the compass of the model.
The interpretation of  the lagrangian density with `pressure' is natural and
leads to the equation of state
$p =  a(\gamma-1)\rho^\gamma$; we expect that the constant $\gamma$ can be
related to $C_v/C_p$. Thus the thermodynamical aspects of the theory are derived
from within the theory itself.

Section 4 is devoted to a study of the case of weak fields. In this limit
our model reduces exactly to the theory developed by Emden [Em] and Eddington
[Ed],
although the equations that define their theory did not originally come from
an action principle. We shall raise some questions about their
choice of boundary conditions.

Eddington dismisses the possibility of gaseous distributions extending to
infinity because they exhibit a singularity at the center.  Chandrasekhar
and others have proposed to accept such singularities as a manifestation of a
local breakdown of the gas laws. Eddington's real problem was that, being
prevented (by the singularity) from formulating boundary conditions at the
origin, he was also unable to fix the solution by
imposing asymptotic conditions, this because his equations do not
involve the gravitational potential $\nu$ itself, but only its derivative.  
In the approximation of weak fields, the only innovation that
our theory brings with it is that it repairs this particular defect, making it
possible to impose asymptotic boundary conditions and to make precise
predictions for the singularity at the center.

The polytropic index is the number $n$ defined by
$$
\gamma = 1 + {1\over n}.
$$
It has been known, since the early work of Emden [Em] that the value $n = 5$
separates two qualitatively different regimes:
 
\ce{$0<n<5:$ "Gas spheres" of infinite extension.}

\ce{$5<n<\infty:$ Finite stars with boundary.}
 
 Section 5  studies the regime of infinite distributions, with
applications to stellar atmospheres and possibly to gaseous giants such as
Betelgeuze and Capella; we present  results of numerical integration of the
equations that determine the equilibrium in the model.  The
exact solutions differ from those of the weak field approximation to a degree
that probably could not have been predicted. The density is strongly peaked in
the neighbourhood of the Schwarzschild radius; there is a central region with low
density (and lower temperature).  There is also a very small central core where
the Emden  function $p/\rho$ turns negative and where it is believed that the gas
laws break down.

Section 6 deals with the regime of stars with boundary, 
spherical stars with  matter confined to $r < R$. They are  
  characterized, according to Eddington and others, by the fact
that the density reaches zero at $r = R$.
But   this choice of boundary conditions   seems to us to be
dubious, and  certainly not possible in the proposed model, where the
boundary is determined by  matching to the exterior Schwarzschild
metric,
$$
\nu(R)+\lambda(R) = 0,~~  \lambda(R) = \ln(1-2m/R).
$$
The first condition determines $R$ and the second one determines $m$. Eddington
has only the second equation at his disposal.  We apply this boundary condition
and examine the solutions near the center of the star, to discover a critical
relation between mass and radius.

In this paper 
 the effect of radiation pressure (important for masses larger than that of
the sun), as well as other necessary refinements, have not been taken into
account. For this reason we do not discuss applications to real stars, except
for a tentative discussion of neutron cores at the end.

In this paper the units are such that $  G  = c = 1$. 
\b\b

\no{\steptwo 2. A matter model.}

A standard model for irrotational flow describes the state
of a continuous distribution of matter in terms of a density $\rho$ and a
velocity potential $\Phi$, the velocity   being the negative
gradient of $\Phi$. The equation of motion for the velocity (the hydrostatic
condition),  and the continuity equation for the current, are both derived from
the variational principle with action ([FW] p. 304)
$$
 \int d^3x \Big(\rho\dot \Phi -{\rho\over 2}\vec v^2 - V[\rho]\Big),  
 $$
where the internal energy density $V$ often depends only on $\rho$ (isentropic
case) and is determined by the equation of state. The expression that is most
commonly used for $V$ contains a term linear in $\rho$ that represents the
external force and in addition a term
$ 
\rho S[\rho],
$ 
where $S$ is the internal energy per unit of mass.  A relativistic version of
this theory has two scalar fields,
$\rho$ and
$\psi$, and the action
$$
A = \int d^4x\sqrt{-g}\Big({\rho\over 2} g^{\mu\nu} \psi_{,\mu}
\psi_{,\nu} - W\Big) =:  \int d^4x\sqrt{-g}~{\cal L}~.\eqno(2.1)
$$
The nonrelativistic theory in flat space is recovered with
$$ 
\psi = c^2t + \Phi,~~ W = {c^2 \over 2}\rho + V.
$$ 
(Everywhere but here we have set the velocity of light equal to 1.)

The vector field $\Psi$ defined by
$$
\Psi^\alpha =  g^{\alpha\beta}\psi_{,\beta}
$$
can be interpreted as a four-dimensional flow velocity. The physical
three-velocity $ -\vec \Psi/\Psi^t$ is invariant under local scaling; the field
$\rho$ acts as a gauge fixing field, eliminating the unphysical, fourth degree of
freedom.  The traditional approach is to fix the normalization by hand. This
amounts to renormalizing  the density and one has to show that
this renormalization qualifies the density for the role assigned to it in
thermodynamics, specifically in the equation of state; once an equation of state
is imposed one is no longer doing pure phenomenology. We are not aware of any
reference where this question is brought up, let alone settled.   

The energy momentum tensor is
$$
T_{\mu\nu} = \rho\,\psi_{,\mu}\psi_{,\nu}- g_{\mu\nu}~{\cal L}~.\eqno(2.2)
$$
In particular, in the metric (1.2),
$$
T_t^t = \e^{-\nu}\rho(\dot \psi)^2 -{\cal L}, ~~ T_r^r = -\e^{-\lambda} 
\rho(\psi')^2 - {\cal L},~~ T_t^r = -\e^{-\lambda}\rho\psi'\dot\psi~. 
$$   
The matter field equations are
$$
{1\over 2}( g^{\mu\nu} \psi_{,\mu}
\psi_{,\nu}-1)
 = (d V/d \rho),~~
\p_\mu\big(\sqrt{-g}\,
 \rho\, g^{\mu\nu}\psi_{,\nu}\big)=  
0. \eqno(2.3)
$$
The first equation fixes the scale; it comes from variation of the field $\rho$.
The second equation comes from variation of the field $\psi$; it is a
conservation law for the current
$\sqrt{-g}\rho g^{\mu\nu}\psi_{,\nu}$.

A common extra ingredient in most works in this area is a conserved quantity
identified as baryon number. In the model this role is taken by
the field $\rho$. In the non-relativistic theory
$\rho$ is the mass density   and
$\rho V$ is the internal energy,  inclusive or not of rest mass. 

A consequence
of (2.3) is that
$$
{\cal L} = \rho{\p W\over \p \rho} - W = \rho{\p V\over \p \rho} - V,
$$
which shows that the lagrangian density ${\cal L}$ is the pressure. ([FW] page
304.)
	In the case that
$dV/d\rho = 0$ the first field equation gives 
$$
g_{\alpha\beta}\Psi^\alpha \Psi^\beta - 1 =
g^{\alpha\beta}\psi_{,\alpha}\psi_{,\beta} -1 = 0  ~~({\rm when~} dV/d\rho = 0 ).
\eqno(2.4)
$$

\no This condition is appropriate only in the case 
that  no forces other than   those that of gravity are present. 
  There will be   discussion of this point in Section 3.

This model of a matter distribution in the
context of General Relativity is an alternative, under
the limitation to irrotational and isentropic flows, to the traditional approach
first proposed by Tolman [T], and the simplest one possible. In Tolman's theory
the energy momentum tensor is expressed in terms of a velocity field,   {\it a
priori} normalized, and two functions interpreted as energy
density and pressure. The pressure corresponds to the function 
${\cal L}$ and the energy density to the function $  g^{00}\rho\dot
\psi^2 -{\cal L}$. Most accounts  of stellar dynamics assume that the
entropy is either constant or irrelevant   ([MTW], page 599)  so that the
thermodynamics is both irrotational and isentropic. An equation of state is
imposed  by hand. The existence of
dynamics in the form of a variational principle is plausibly inferred from
the Bianchi identities via Einstein's equation, but only the Bianchi
identities themselves were invoked.   

The model studied here is not yet fundamental theory, since the choice of the
potential $V[\rho]$ remains open, but it goes a step beyond
Tolman's theory.  Once the potential has been fixed nothing but the imposition
of boundary conditions remains. If the model admits solutions that
account  for observed phenomena then something has been learned concerning the 
types  of matter distribution that are consistent with Einstein's equations and
thus suitable for the framework of General Relativity.

We shall try to find realistic solutions of the system that consists of
Einstein's equations,
$$
G_{\mu\nu} = 8\pi T_{\mu\nu}.
$$
together with the Euler-Lagrange equations (2.3).
The metric will be assumed to have rotational symmetry. We assume that
there are coordinates $t,r,\theta,\phi$ such that the line element takes
the form
 $$
(ds)^2 = \e^\nu(dt)^2 - \e^\lambda(dr)^2  - r^2d\Omega^2,\eqno(2.5)  
$$
with $\nu$ and $\lambda$ depending only on the coordinates $t$ and $r$, and
we ask that
$\rho$ and
$\psi$ also depend only on $t$ and $r$; there are then 4
independent equations,
$$
G_{tt} =  T_{tt},~~ G_{tr} =  T_{tr},~~ G_{rr} = 
T_{rr},\eqno(2.6-8)
$$
and
$$
G_{\theta\theta} =  T_{\theta\theta}.
$$
Because of the Bianchi identities, satisfied identically by the  
Einstein tensor and by virtue of (2.3) by the energy momentum tensor,
the last equation is a consequence of Eq.s (2.6-8) and can be ignored.
\b

\ce{\bf Summary of equations }
 
Einstein's equations,
$$\eqalign{&
G_t^t =  -e^{-\lambda}\Big({-\lambda'\over r}+ {1\over r^2}\Big) +
{1\over r^2} =  8\pi\Big(\e^{-\nu}\rho\dot\psi^2 -{\cal L}\Big),\cr 
& {G_r}^r =  -{\rm e}^{-\lambda }\Big({\nu'\over r} + {1\over r^2}\Big)+
{1\over r^2} =  8\pi\Big(-e^{-\lambda}\rho(\psi')^2 - {\cal L}\Big),\cr
&G_t^r = \e^{-\lambda} {\dot \lambda\over r} =  - 8\pi\e^{-\lambda}\rho
\psi'\dot\psi.
\cr}\eqno(2.9-11) 
$$

Wave equations,
$$\eqalign{&
{1\over 2}\Big(\e^{-\nu}\dot\psi^2-\e^{-\lambda}(\psi')^2-1\Big) =
{d V\over d\rho} , \cr 
\p_t(&\e^{(-\nu+\lambda)/2}r^2\rho\dot \psi
)-(\e^{(\nu-\lambda)/2}r^2\rho\psi')'= 0.\cr}\eqno(2.12-13)
$$
 
The function $\lambda$ is often replaced by the function $M$ defined by
$$
M: = {r\over 2}(1-\e^{-\lambda}),~~ \e^{-\lambda} = 1 - {2M\over
r};\eqno(2.14) 
$$
 then Eq.s (2.9-10) can be written as follows,
$$\eqalign{ 
M' &=  4\pi\, r^2(\e^{-\nu}\rho\dot \psi^2-{\cal L}),\cr
r\e^{-\lambda}\nu' &= 1-\e^{-\lambda} + 8\pi\, r^2(\e^{-\lambda}\rho \psi'^2 +
{\cal L}). 
\cr}\eqno(2.15-16)
$$
The two equations can be combined to yield
$$
 (\nu + \lambda)'  = 8\pi\, r\,\e^\lambda\rho\,(\e^{-\nu} \dot\psi^2 +
\e^{-\lambda}\psi'^2).\eqno(2.17)
$$
 
\no{\bf Remark.}  Comparing our equations with those that appear
in the literature, from Tolman [T] onwards, we
see that what is normally called ``energy density"  
is identified with
$ 
  \e^{-\nu}\dot\psi^2\rho -   {\cal L}.
$ 
 A very important aspect of
equations (2.9-13) is the appearance of the function   $\nu$. The equations of
the traditional approach contain the derivative of this function but not the
function itself; it is therefore meaningless to impose asymptotic boundary
conditions.   Eddington [Ed], and modern authors as well, fix the boundary of a
star at the first zero of density and pressure. There is no  matching to an
outside or asymptotic Schwarzschild metric, and the ``mass'' of the star is
instead defined by the value of the function $M$ introduced in (2.14). This is
a major shortcoming of the traditional approach to relativistic stellar
dynamics. 
 The
traditional  pressure  
corresponds to 
$ 
{\cal L} + \e^{-\lambda}\rho \psi'^2.
$ 
In the static case $\psi' = 0$ and there is full agreement since, as
already pointed out,  our ${\cal L}$ is interpreted as pressure in the
nonrelativistic limit. But here too, a quantity normally interpreted
entirely in terms of matter is found generally to depend on the metric as
well.  
\b
 \b

\no{\steptwo 3. The static limit} 

 In the context of constructing an
`interior solution', to be joined to an exterior Schwarzschild metric, it
is common to define ``static" by fixing the four-vector field, setting the
space components to zero and normalizing  $U_t$ as in (2.4), thus
$g^{tt}U_t^2 = 1$.   We define the  static limit by
$$
\dot\psi = 1,~~\dot \rho = \dot p = \dot \nu = \dot \lambda = 0,
$$
and Eq.(2.11) then does not allow any reasonable alternative to setting
$\psi' = 0$. Conversely, if we admit that `static' implies no flow,
$\psi' = 0$, then $\dot\psi$ is independent of position and then the
equations demand that it be a constant. Taking $\dot\psi =
1$,  in agreement with the nonrelativistic limit, and the only value
possible in Minkowski space, is convenient.   The full set of
equations in the static case is thus as follows,
$$ 
M' =  4\pi\, r^2 (\e^{-\nu} \rho - p),   ~~~~
  (\lambda + \nu)' = 8\pi\, r\, \e^{\lambda-\nu}\rho,~~~{1\over 2}(\e^{-\nu} -
1) = {dV\over d\rho},\eqno( 3.1 -3) 
$$
with
$$ 
 p := {\cal L} = \rho{d V\over
 d\rho} - V  . 
$$ 
The first two equations agree with the textbooks if we identify
$\e^{-\nu}\rho-p$ with the energy density  (denoted $\hat \rho$).
Derivation of the third equation with respect to $r$ gives
$ (\e^{-\nu})'/2 =  p'/\rho$, which is known as the hydrostatic equation
([Ed], page 79); this equation is a consequence of the (contracted) Bianchi
identities and is used by Chandrasekhar and by Oppenheimer and Volkoff  as
well [OV]. Conversely, Eq. (3.3) and the Bianchi identities makes the
other wave equation (the conservation law) redundant, so that the only thing
that is new in our treatment of the static case is the fixing of an integration
constant. 

Since this point is important we insist on it.  
Eq.(3.3) is equivalent to the pair
$$\eqalign{&
(1)~~~
 {1\over 2}(\e^{-\nu})' = ({dV\over d\rho})',\cr
&
(2)~~~{1\over 2}(\e^{-\nu} -1) = {dV\over d\rho} ~{\rm at ~a~ point}.  
\cr}
$$
The quantity $({dV\over d\rho})'$ is easily seen to be the same as $p'/\rho$,
and the first equation is then recognized as the hydrostatic condition.
There remains the second statement that,  as we said, fixes an integration
constant. This equation is new, there is no precedent in the traditional
theory. We shall see that it is a positive development. 

An important corollary is that matching the solution to an exterior
Schwarzschild metric constitutes a complete set of boundary conditions, which is
not true in the traditional setting.  Another application follows.

In the case of an polytropic equation of state Eq.(3.3) takes the form
$$
{p\over \rho} = {\phi\over  n+1 },
$$
where $\phi $ is the gravitational potential (defined by $\e^\nu = 1-2\phi$) and
$n$ is the polytropic index. Defining the temperature by the gas law, $pv =
RT/\mu$, where
$\mu$ is the atomic weight, we convert this to
$$
T = {\mu\over R}{\phi\over n+1}.
$$
The standard approach yields the same result, but modulo an additive constant.
(Compare [L], p. 268-271.)  For a polytrope with $n = 3, ~\mu = .5$, with a 
sharp boundary where pressure and density drop abruptly to zero, and with the
ratio mass/radius  equal to that of the Sun,  this formula yields  a
temperature at the surface,
$$
T = {1\over 8}{1\over 8.3\times 10^7}{(2\times 10^{33})(6.6\times
10^{-8}) 
\over 7\times 10^{10}} =  2.7\times 10^7 ~{\rm degrees~Kelvin.} 
$$

\b
       
\ce{\bf Empty space } 

We assume that empty space can be viewed as a
limiting case of equilibrium. Though it
is difficult to interpret the velocity of flow when no matter is present, we set
$\psi' = 0$.  If it should happen that $dV/d\rho = 0$ (in empty space),
then we get the result that the gravitational potential $-\nu/2$ vanishes.  Is
this a  reasonable conclusion?
 
  Let us remember that the idea of freezing the value of
 $g_{\mu\nu}U^\mu U^\nu$ at unity has its origin in the equation
for geodesic motion of a test particle; it is in that case a constant
of the motion that can  be fixed. So doing we also eliminate an extra,
uninterpretable variable, $U^t$. (It is the relativistic version
of the condition that $E-H = 0$.)   We are thus talking about a  test
particle in the gravitational field, with 3-velocity equal to zero  (since
we are still dealing with equilibrium). But this is absurd, except in
Minkowski space. Indeed, if matter is at rest ($\psi' = 0$) in any other
gravitational field then there must be a force  present, to balance the
gravitational  pull on the particle; then the motion is not along a geodesic and
then we do not expect that
$g_{\mu\nu}U^\mu U^\nu$ should be a constant of the motion. That is exactly
what Eq.(3.3) tells us. The quantity
$dV/d\rho$ represents the balancing force and it is zero only when the
gravitational potential is zero. The aptness of the equation is indicated by
the fact that taking the derivative with respect to $r$  
  one obtains the hydrostatic condition.

Thus we regard empty space as a limit of a
space filled with matter. This will turn out to be a clue to understanding
the Schwarzschild solution.

\b
\ce{\bf Equation of state}

An equation of
state, for isentropic processes, is a relation between density and pressure. This
brings the difficult question of what is the correct definition of the
density field. In the limit of weak fields there is no doubt that  
$\sqrt{-g}\,\rho-p$ is the density of mass, whose integral determines the mass
parameter in the asymptotic Schwarzschild metric. But it is not clear that 
the density of mass is what should appear in the equation of state. It appears
that there is no general answer, and no clear consensus on this topic. We have
already quoted, in the introduction, an instance where an equation of state was
used with different choices of density. Being unable to decide this question,
one can only say that the equation of state selected may be appropriate in some
cases. As a practical matter, it turns out that different interpretations of the
``density" in the polytropic equation of
 state makes very little difference in the cases that we have studied. This also
renders moot the question of whether the traditional normalization of the
velocity four-vector leads to the correct definition of the density to be used
in the equation of state; nevertheless, there is an important matter of
principle.

Our point of view is a little different. We are going to take a very simple
choice of the functional $V$, thus defining the theory. Then we let the theory
speak for itself. The interpretation of a theory is in its consequences. In fact
it will give us plenty of scope for a thermodynamical interpretation.

 In this paper we shall study the consequences of taking 
$$
V =  a\rho^\gamma,~~{\rm thus}~~p = a(\gamma-1)\rho^\gamma,~~ a,\gamma ~{\rm
constants},~~ \gamma > 1.\eqno(3.4)
$$
\b
\ce{\bf Asymptotic behaviour}
  Let us
consider a matter distribution that extends to great distances and suppose
that, as
$r$ tends to infinity, $\rho = br^{-3-\epsilon}$ with $b,\epsilon$
constant and $\epsilon >0$ to make the density integrable at infinity. Thus
$$
\rho \sim r^{-3-\epsilon}~~\Rightarrow\hskip.3cm\matrix{M' \sim 
(r\lambda)' 
 \sim r^{-1-\epsilon} \cr\cr \lambda' + \nu' \sim r^{-2-\epsilon} \cr\cr
 \nu \sim  dV/ d\rho 
\sim r^{(3 +\epsilon)(1-\gamma)}\cr},\hskip1cm\epsilon > 0.\eqno(3.5) 
$$
For the time being we shall consider only the case when this equation of
state applies for all $r$ larger some limit. We are thus dealing with a
stellar atmosphere, though there may be nothing else, as in the case of
the young giants. 
 
Inspecting (3.5), we are  at first  
tempted to conclude that $\lambda$ and $\nu$ must fall off faster than
$1/r$,
which would exclude solutions that behave like the Schwarzschild  metric at
large distances.  In this case 
$1+\epsilon =  (3 +\epsilon)(\gamma-1)$, or $\gamma = (2\epsilon +
4)/(\epsilon + 3) > 4/3$. The  way to
circumvent this is to assume that $\lambda \sim r^{-1}$, for in that case
the leading term in $\lambda$ makes no contribution to $(
r\lambda)'$. This term also has to make no contribution to $\lambda + \nu$,
so we can assert that there is a constant $m$ such that
$$ 
\lambda \sim {2m\over r},~~ \nu \sim -{2m\over r},~~(m = {\rm constant}).
$$ 
This implies that $(3 +\epsilon)(\gamma-1) = 1$, or
$$
\gamma = 1 + {1\over 3 + \epsilon},~~  1 < \gamma < {4\over 3},
$$
which is   acceptable; the upper limit is slightly less than the value
for an ideal,  diatomic gas.\break 
 \b
\ce{\bf  \hskip-2cm Result for polytropic stars  without boundary}
  (1)  If the  density of the stellar atmosphere falls off fast
enough   

to be   integrable at infinity,      
 then at large distances the metric

approaches   the Schwarzschild metric. 
\b 
  (2) The static boundary conditions, in the case of the
isentropic 

 equation of state, specifies the leading terms
for large $r$ as 
$$
\rho \approx b r^{1\over 1-\gamma},~~ p \approx a(1-\gamma)b^\gamma
r^{\gamma\over 1-\gamma}.
$$
 
   For the metric, if $0<\epsilon<1$, the three leading terms for large $r$
are
$$
g_{tt} =  e^{ \nu} \approx  1 -{2m\over r} + {8\pi b\over \epsilon(1 +
\epsilon)} r^{-1-\epsilon},~~ g_{rr} = \e^\lambda \approx 1+
{2m\over r} - {8\pi b\over \epsilon} r^{-1-\epsilon} . 
$$

(3) The constant  $\gamma$ has to lie in the interval $1 < \gamma < 4/3$. 
\b\
The assumed constancy of $\gamma$ is of course an oversimplification of the
situation in real stars, see [M] , [Fo] or [KW] page  175. 
\b

\ce{\bf The Schwarzschild metric } We ask if there is a singular limiting
matter distribution in which the metric becomes exactly Schwarzschild.
Assume that
$$
\e^{\nu} = 1 -  {2m\over r},~~ m = {\rm constant},
$$
and then Eq. (3.3) gives us (precisely)
$$
{m\over r- 2m} = a\gamma \rho^{\gamma-1},~~ \rho = b(r-2m)^{1\over
1-\gamma },~~ b = ({m\over a\gamma})^{1\over \gamma-1},
$$
which makes sense for $r > 2m$.
Eq.s (3.1-2) now read
$$\eqalign{
M' &=  4\pi\,b \,r^2  \big( r +   {1-\gamma\over \gamma}m  
   \big)(r-2m)^{\gamma\over 1-\gamma}  ,\cr &(\nu + \lambda)'
= 8\pi\, b\,r\e^{\lambda-\nu}(r-2m)^{ 1 \over
1-\gamma}.
\cr}
$$
Both expressions must tend to zero in empty space, in the limit that we are
looking for, which requires that $\gamma$
tend  to 1 and that
$a \geq m$. In the limiting case when $a \approx m$, and $\gamma$ is very
close to 1,
$$\eqalign{
&M'  = 4\pi\,b \,r^3   (r-2m)^{1\over 1-\gamma} 
,\cr &(\nu + \lambda)' =  8\pi\,b\,r\e^{\lambda-\nu} (r-2m)^{ \gamma \over
1-\gamma}.
\cr}
$$
the pressure and the density
both tend to zero except at $r = 2m$. But $dV/d\rho$ does not tend to
zero; the gradient $(dV/d\rho)' = p'/\rho$ represents the force per
unit volume needed to resist the pull of gravity even in the limit of
vanishing density.   

These conclusions are supported by numerical calculations presented in
Section 5.

  An analogy may serve to support 
this interpretation of the Schwarzschild metric.  Consider Poisson's
equation in the case of a source localized at the origin. 
One solves Laplace's equation in empty space ($r\neq 0$) and finds
that the solution is const.$/r$, which is singular at the origin. A fine
argument, using Gauss' theorem, shows that there is a non-vanishing
source, given by a   distribution concentrated at the origin. An
alternative would be to consider a sequence of continuous distributions of
matter, converging towards a
$\delta$-function distribution. In General Relativity one solves Einstein's
equation for empty space. The solution is ``singular" at the origin and at
$r = 2m$. The presence of a source is manifest, but where is it localized?
The region inside the Schwarzschild radius is unphysical [D], so to localize
the source at the origin begs the question [Fr]. The singularity at $r = 2m$
turns out to be removable by a coordinate transformation and the naive
conclusion is that the source is nowhere!  (Skeptical,
  we examined the possibility that the Einstein tensor,
evaluated for the Schwarzschild metric, might contain a distribution of
the form $\delta(r-2m)$. It does not, but there is more to this question.)

What the model shows  is that a different distribution, not a
$\delta$-function but nevertheless concentrated to a  high
degree at the horizon, yields a metric that is  close to the
exterior Schwarzschild metric except in a region near the Schwarzschild. We
suggest that this type of localized distribution may be what is relevant in the
context of Einstein's equations. In that case we could affirm that the source of
the Schwarzschild metric is localized at the horizon.

By including a dynamical model of the source of the gravitational
field, we impose additional constraints on the theory and are led to a solution
of  the problem of localizing the source of the Schwarzschild metric field in a
physically meaningful region. Without matter or boundary conditions  we
cannot hope to find the correct interpretation.

Consider the case of a spherically symmetric distribution of matter, in the
form of a spherical shell centered at the origin, and of negligible
thickness. In the newtonian theory of gravitation this gives rise to a
gravitational potential outside the shell, but it has no effect on the
inside! Therefore, if one adopts the hypothesis that the Schwarzschild
metric is due to a concentration of mass at the horizon, then it is absurd
to pretend that  the metric field inside is obtained by analytic
continuation. Much more natural to suppose that the metric is
discontinuous at the horizon; for example,
$$
\e^\nu = 1-\theta(r-2m){2m\over r} =\e^{-\lambda},
$$
where $\theta$ is the unit step function, and the non vanishing components
of the Einstein tensor are
$$
G_t^t = G_r^r = {1\over 2m}\delta(r-2m),~~ G_\theta^\theta =
G_\phi^\phi = {1\over 2}\delta'(r-2m).
$$

Incidentally, the odds against the mythical traveller
 penetrating the horizon during his life time needs to be
re-evaluated!
\b\b

\no {\steptwo 4. Weak fields}

  At
low densities the metric functions $\lambda,\nu$ will be small compared to
unity, and Eq.s (3.1-3) take the form 
$$
 M' = 4\pi\,(r\lambda)'  =  r^2(\rho-p),~~ (\lambda + \nu)' =
8\pi\, r\rho, ~~\nu = -2a\gamma\rho^{\gamma-1}.\eqno(4.1)
$$
   It follows
from the third equation  that $p  = -{\gamma-1\over 2\gamma}\nu\rho$, so
that the contribution of the pressure to the first equation must be
omitted in the weak field approximation. 
 
 In the unphysical case that
$\gamma = 2$ the equations become linear,
$$
\rho = -\nu/4a,~~ \lambda = r\nu',~~  \lambda' + {1\over r}\lambda
= -  2\pi\,r\nu/ a.
$$ 
Eliminating $\lambda$ we get the radial Schroedinger equation for a free
particle with angular momentum zero,
$$
 \nu'' +  {2 \over r}\nu' + {2\pi\over  a}\nu = 0,
$$
One solution  is
$$
\nu = c\, {1\over r}\sin(kr),~~ k^2 = {4\pi\over a},~~  ~~c = {\rm
constant}.
$$  
However, the value 2 for $\gamma$, implying $\epsilon = -2$ is not
an option for an infinitely extended distribution.

Let $\gamma = 1 + 1/n, ~ n >1$. Continue to ignore the
contribution of the pressure. The equations are now, with
$f^n =
 \rho$,    
$$
2M' = \lambda + r\lambda'  = 8\pi\, r^2 f^n ,~~ (\lambda + \nu)' =
8\pi\, rf^n, ~~\nu = -2a\gamma f,\eqno(4.2)
$$ 
We reduce this to the nonlinear Schroedinger equation (Emden's equation)
$$
\eqalign{&
f'' + {2\over r} f' + k^2   \,f^n,\cr
& \nu = -2a\gamma f,~~ \lambda = r\nu', ~k^2 = 4\pi/a\gamma.\cr}\eqno(4.3)   
$$

 When $n = 5$ there are  exact solutions  ([Ed] page 89),
$$
f = \alpha(1+\beta^2 r^2)^{-1/2}, ~~ \beta^2 = {8\pi\over 6a\gamma
}\alpha^4.
$$ 
Thus,
$$\eqalign{
\rho & = \alpha^5(1+\beta^2 r^2)^{-5/2}, ~~ p = {a\alpha^6\over
5} (1+\beta^2 r^2)^{-3}\cr
 &-\nu/2 =  a\gamma\alpha(1+\beta^2 r^2)^{-1/2},~~ \lambda =
r\nu'. 
\cr}
$$
Asymptotically, $m := \lim(-r\nu/2) =  a\gamma\alpha/\beta 
   $ is the gravitational mass.
The integrated density is
$$
4\pi\int_0^\infty r^2 \rho dr =4\pi \alpha^5\beta^{-3 }\int z^2(1 +
z^2)^{-5/2}dz = (8\pi/6) \alpha^5\beta^{-3 }\int  (1 +
z^2)^{-3/2}dz$$
The last integral equals $\int_o^\infty dx/\cosh x = 1$ and finally
$$
4\pi\int_0^\infty r^2 \rho(r) dr   = a\gamma\alpha/\beta =m, 
$$
as expected. In this context  $\rho$ is the density of
mass.   Solutions obtained numerically for
$n > 5$ are qualitatively similar. A limiting case of infinite $n$,
or $\gamma = 1$ is discussed in [Ed], page 89. The case of $n< 5$ will be
taken up in Section 6. 
\b 
\no{\bf Remark 1.}  The relation $\nu = -2a\gamma f$ tells us that the
potential and the Emden function are of the same order of magnitude. The
hydrostatic equation used by Eddington has the same implication. Therefore, if
$n\geq 2$, then consistent application of  the weak field hypothesis would
require us to drop the term
$k^2f^n$. Nevertheless, we shall refer to these equations as the weak field
approximation, since they are the equations proposed and studied by Eddington
and generally considered to be valid in the case of weak gravitational fields.
See [Ed], page 80.
\b
\no{\bf Remark 2.} We shall always require our solutions to be
asymptotically Schwarzschild. Besides identifying the gravitational mass with the
asymptotic limit of $-r\nu(r)$ it requires that $r \nu(r) + r\lambda(r)$ tend to
zero. This is a departure from the work of Eddington. In his work the equation
$\nu = -2a\gamma f$ (coming from Eq. (2.12)) is replaced by the hydrostatic
equation
$\nu' = -2a\gamma f'$. Integration introduces an undetermined, additive constant
and matching the metric to Schwarzschild becomes less restrictive.  Eddington's
boundary conditions are $f(0)$ finite and $f'(0) = 0$.

\b\ve

\no{\steptwo 5. Numerical integration for an infinite star}

  Numerical
solutions of the system (3.1-3) that govern static configurations were obtained
with the help of Mathematica.
\b
\ce{\bf Parameters}

A scale transformation, by which the coordinate $r$ is replaced by a
new coordinate
$$
x = \sqrt{8\pi a^{-n}} \,r
$$
 allows, without
essential loss of generality, to set $a = 1$  in $V[\rho] = a\rho^\gamma$.
(Recall that $p$ and $\rho$ are scalar fields, not densities). The constant
$\gamma$ is expressed as
$$
\gamma = 1 + {1\over n}.
$$
The choice of the value of $n$ is the only freedom that we have
explored with respect to the thermodynamic properties of  matter, except
for boundary conditions. In this section we restrict $n$ to the regime of
infinite stars, $n>5$.
\b
\ce{\bf Boundary conditions}

It was seen that the density of a star without boundary is integrable at
infinity if and only if the metric is asymptotically Schwarzschild. This result
reflects the fact that the equations of the model, unlike the traditional ones,
are not invariant under a constant shift of the function $\nu$. It allows us to
fix a solution in terms of asymptotic boundary conditions, while all previous
work has applied conditions at the center of the star, a region about which
little is known in advance. (A notable exception is the paper [C].) Thus
Eddington supposes that the metric is well behaved at the origin, more precisely
that
$\nu'(0) = 0$, and concludes that all stars without boundary have infinite mass.
(  [Ed], page 88.) It turns out that, in the traditional theory, the solutions
that are regular at the origin do not have finite mass; one has to choose. (We
are  under the restriction $n >5$.)

It will be useful to compare our results with those of the weak field
approximation. 
The equations used by Eddington [Ed]  and in all subsequent
work are the same as those of our model, except for two modifications.
\b
1. Our equation (2.12),
$$
{1\over 2}(\e^{-\nu}-1) = dV/d\rho
$$
differs from the usual hydrostatic condition only in the fact that it determines
the integration constant, which gives a meaning to the value of the function
$\nu$ and not just its derivative. This affects only the boundary conditions. It
gives us the possibility of fixing a unique solution by the imposition of
asymptotic boundary conditions. The usual  practice is to
integrate outwards from the origin, this implies regularity near the
origin and no solution of this type is integrable at infinity, as Eddington
remarks. ([Ed, page 88.)

2. To relate one system of equations to the other we have the dictionary
$p\mapsto \hat p,$ ~ $\e^{-\nu}  \rho \mapsto \hat\rho + \hat p$, so that the
equation of state, though formally the same in the two theories, does not
involve the same variables.   
\b 
Solutions are completely fixed by the choice of $\lambda(x_0)$ and
$\nu(x_0)$ at some point $x_0$.  By trial it is found that
$$
\nu(25) = -.1,~~\lambda(25) = .1
$$
generates solutions for $\nu,\lambda,\rho$ and $p$ that fall off
uniformly at large $x$. In the first place, the point chosen is more
than 5 times the  value of $x$ where bumps appear, so that $(\nu +
\lambda)(25)$ is close to the asymptotic value zero. In the second
place, we expect to find solutions for which $x\nu(x)$ and $x\lambda(x)$
tend to finite asymptotic values, which forces the value of
$\lambda(25)$ to agree with that of $-\nu(25)$ to within one part in a
thousand.

The number $-25 \nu(25) = 2.5$ is close to the asymptotic value of
$-x\nu(x)$ at infinity; that is, to the parameter $2m$ in the
asymptotic field $g_{00} = \e^\nu = 1 - 2m/x.$   The only free parameter
is thus the gravitational mass $m$. We have not explored a wide range of
values of this parameter but some sampling suggests that things do not
change qualitatively. Thus Fig.s 1a-d show that a factor of 4, or even a factor
of 40, in the value of
$2m$ leaves the basic shapes intact.
\b

\ce{\bf Results}

1. In a first study we varied the parameter $n$ over the set
$\{6,8,10,20,50,100,1000\}$.
In the case  $n = 6$, Mathematica returns reliable answers over the range
$ 
 10^{-300}\leq x \leq 100
\, 000.
$ 
As $n$ grows the permissible range of $x$ contracts at the upper end, due
to strong cancellations in the intermediary stages of the
calculation, but it does not fall below  8 000.  

The metric is well behaved near the origin, both $ g^{tt}$ and $g_{rr}$
approaching zero linearly with $x$. However, the Emden function  
is negative in a very small region near  the center, as shown in the
table.

Results for $n = 6$ are presented in Table 1  and illustrated in Fig.s 1a-f.
The most striking aspect is the strong concentration of the density near the
Schwarzschild radius. Since this is the quantity that sets the scale one
should  not be very surprised.

 The functions $f,\rho$ and
$p$ take positive values, except very close to the center  where the Emden
function $f$ turns negative.  The existence of a hole, with vanishing or
negative pressure, needs  interpretation, but the suggestion that the highest
densities (and temperatures) occur at a distance from the center 
  is, at least, fascinating.

  \vskip1.1in
\def\picture #1 by #2 (#3){
  \vbox to #2{
    \hrule width #1 height 0pt depth 0pt
    \vfill
    \special{picture #3} 
    }
  }
\def\scaledpicture #1 by #2 (#3 scaled #4){{
  \dimen0=#1 \dimen1=#2
  \divide\dimen0 by 1000 \multiply\dimen0 by #4
  \divide\dimen1 by 1000 \multiply\dimen1 by #4
  \picture \dimen0 by \dimen1 (#3 scaled #4)}
  }

\parindent=1pc

\vskip-3cm
\epsfxsize.8\hsize
\centerline{\epsfbox{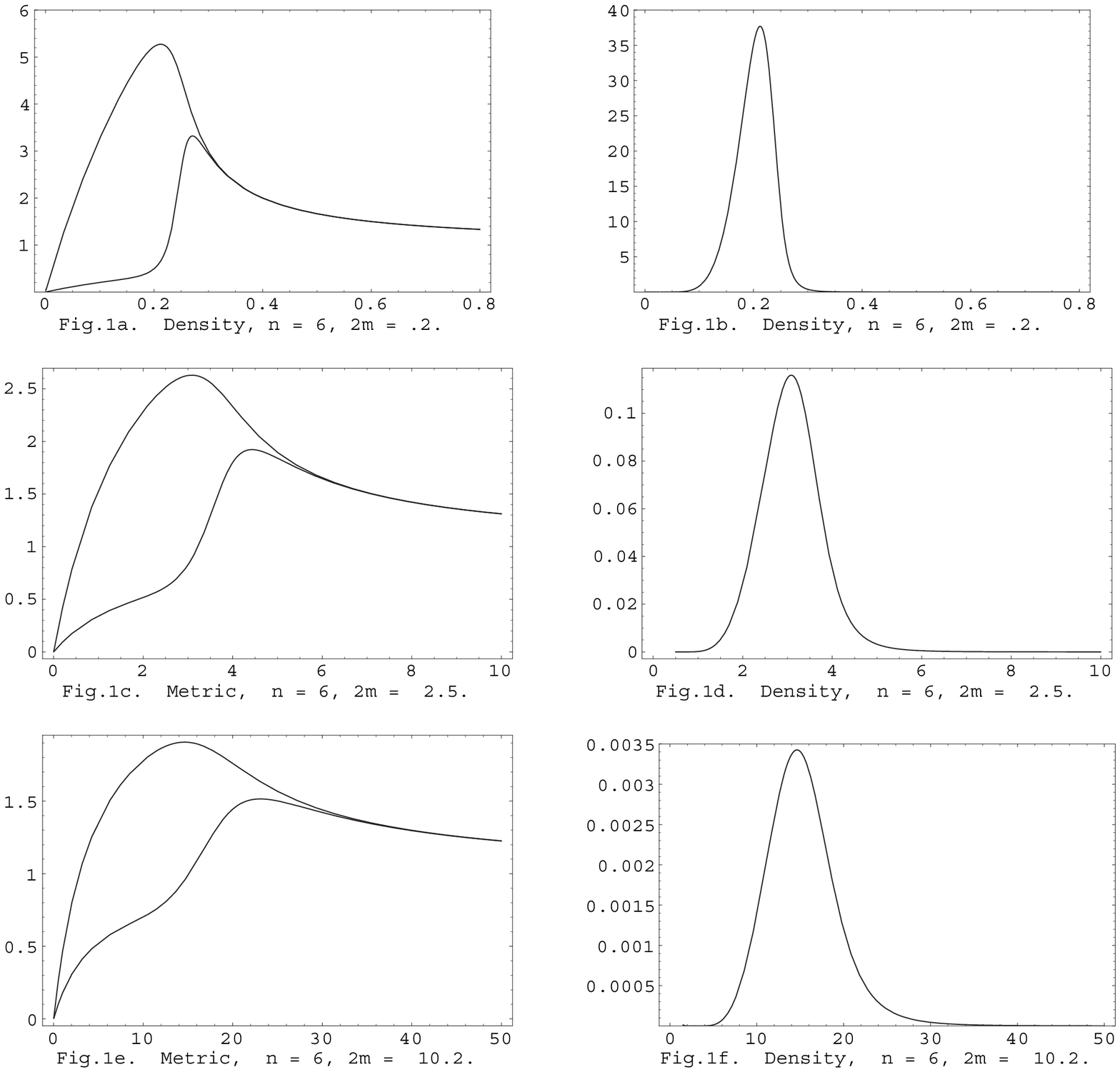}}
\vskip-1cm
\vskip1cm
\no{\it Fig 1. On the left, $\e^{-\nu}$ on top and $\e^\lambda$ below. On the right,
the density. Parameters $n = 6$, $2m$ = .2,~ 2.5 or 10.2  as indicated.

.}

 The functions $f,\rho$ and
$p$ take positive values, except very close to the center  where the Emden
function $f$ turns negative.  The existence of a hole, with vanishing or
negative pressure, needs  interpretation, but the suggestion that the highest
densities (and temperatures) occur at a distance from the center 
  is, at least, fascinating.

  \def\s{ \scriptstyle}
{\stephalf  

\b
 
\ce{\bf Table 1, $n$ = 6, 2$m$ = 2.38}

\settabs  \+   &x   ~~~~~& 0.01 ~~~~~~~~       &
 ~1~~~~~~~~~~~~~~  &3~~~~~~~~~~  ~  & 3.2~~~~~~~~ ~     & 4.43~~~~~~~~  
&~~~~~~~~~~~~~ 10&~~~~~~~~~~~~~~~~~~ &100~~~~~~~~~~~~~~~&1000\cr

\+ & x  & 0.01 &   1 & 3 & 3.2 & 4.43&10&100&1000&100 000\cr

\+&$\s\nu$  & 3.7618 &  
-.43002 & -.96572 & -.96483 & -.75010 & -.27165&  -.02408& -.002381&  
-.00002246 \cr

\+ &$\s  \lambda$ & -5.269  &   -1.0731 & -.19000 & -.02177  &
 .65367&  .27148 &   .02408&  .002382& 
.00002383\cr

\+& $e^{-\nu}$ & .02324 &  1.53729& 2.62669& 2.62435& 2.1172& 1.3121 &  1.0244&
1.00238& 1.00002 \cr
 
\+&$ e^{ \lambda }$ & .0051699&   .341964& .826958& .978462&
1.9226& 1.3119 &   1.0244& 1.00238&   1.00002\cr 

\+&$\s\rho$&  .0053810 &   .0001491& .114806& .113819& .01205 & 5.73$ \s\times
10^{-6}$ &  1.30 
$\s\times 10^{- 12}$ & 1.14$\s\times 10^{-18}$&  
7.94$\s\times 10^{-31}$\cr  

\+& p& -.0003754 &  5.72$   \s\times    10^{-6}$& .013340&
.013206& .0009616 & 1.28$\s\times 10^{-7}$&   2.26$\s\times10^{-15}$&
1.93$\s\times 10^{-22}$&  1.27$\s\times 10^{-36}$\cr 

\+&f& -.41861 &   .23027& .69715 & .69615& .47881&.13377&   .01044  &
.001022&   9.63$\s\times 10^{-6}$\cr}

 \b

2. Table 2a shows a comparison of results of the model with the ``weak field
approximation" (See Remark 1 at the end of Section 4.),  with the same boundary
conditions and with the same gravitational mass,
2$m$ = 2.38, for the case $n = 6$.  
  There is close agreement for $x > 10$, very close for $x > 100$. 
\b
{\stephalf
\ce{\bf Table 2a. Values of   $ -\nu  ,~ n = 6.~2m = 2.38.$}
\settabs  \+~~~~~~~~ &~~~~~ ~~   & x   ~~~~~~~~~& 0.01 ~~~~~   &
 1~~~~~~~~~~~~~ & 3~~~~~~~~~  & 3.2~~~~~~~~  &
4.43~~~~~~~~~  & 10~~~~~~~~& 100~~~~~~ & 1000~~~~~~ & 100 000~~~~  &\cr

\+   &&  x  & 0.01 &   1 & 3 & 3.2 & 4.43& 10 & 100 & 1000 &   100 000\cr

\+&& Weak &  -21.36 &   2.384 & .8318 &
.78006 &  .563035& .2500 &   .02500 & .002500 &  .00002402 \cr 

\+ &-$\s\nu(x) $&Model&   -3.762&  .4300 &  .9657 &  .9648 & 
.750103
&    .2716 &   .02408&  .002381&  
 .00002246 \cr
\+ & &Schwarzschild, -ln$\s(1-2.38/x)$ & &&  1.577 &1.362 &  .7706& .2718  & 
.02401 & .002383 &   .0000238\cr

\b}
 \no  The inner region  is illustrated in Fig.2a-d, where the ordinate is  
$\e^{-\nu}$. 

\vskip1.5in
\def\picture #1 by #2 (#3){
  \vbox to #2{
    \hrule width #1 height 0pt depth 0pt
    \vfill
    \special{picture #3} 
    }
  }
\def\scaledpicture #1 by #2 (#3 scaled #4){{
  \dimen0=#1 \dimen1=#2
  \divide\dimen0 by 1000 \multiply\dimen0 by #4
  \divide\dimen1 by 1000 \multiply\dimen1 by #4
  \picture \dimen0 by \dimen1 (#3 scaled #4)}
  }

\parindent=1pc

\vskip-3cm
\epsfxsize.8\hsize
\centerline{\epsfbox{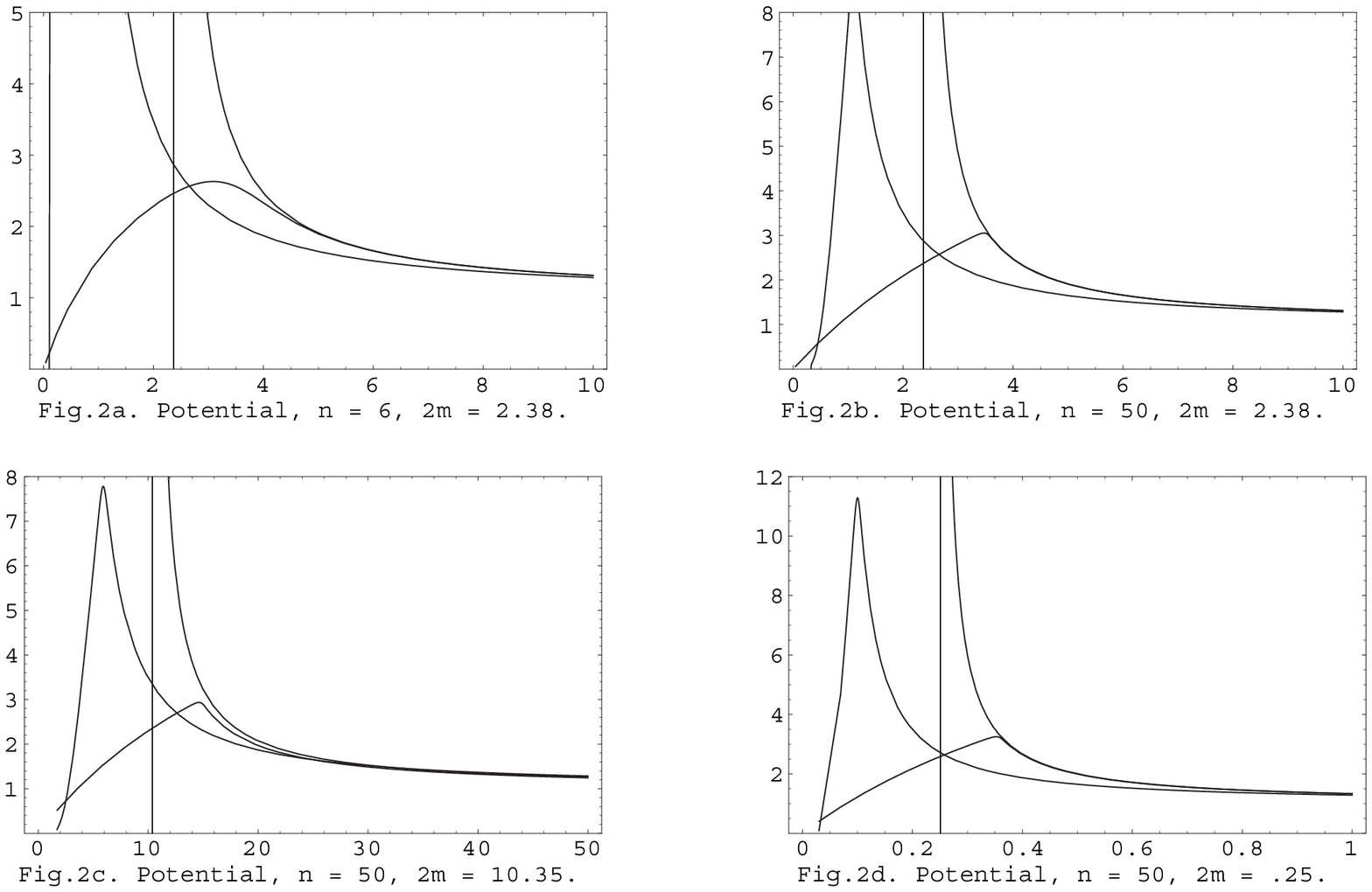}}
\vskip-1cm

\vskip1cm
\no{\it Fig.2. Comparison between the model, the ``weak field approximation" and the\break
Schwarzschild metric in 4 cases. The hyperbola is the   Schwarzschild metric,
$g^{tt} = g_{rr}$. The lower curve is $g^{00}$ in the model. The upper curve,
with the sharp peak inside the Schwarzschild radius, is the same function in the
``weak field approximation".  
}

\b

  Table 2b compares the densities in the model
with the densities of the weak field approximation. The remarkable contrast
between the model and the weak field approximation must have some importance
for our understanding of what is going on near the centers of some stars.

\b
{{\stephalf
\ce{\bf Table 2b. Values of $\rho,~ n = 6$, 2m = 2.38.}

\settabs  \+&      &  ~~~~~~~~    & x ~~~~~~~   & 0.01
~~~~~~    & 1~~~~~~~~~~  & 3~~~~~~~~~    & 3.2~~~~~~~~  &
4.43~~~~~~~   &   10~~ ~~~~~~~~~~~  & ~~~~~~~~~  100~~~~ &~~~~~~~ 100 000 &\cr

\+ &&&x  &  0.1 & 1 & 3 & 3.2 & 4.43& 10 &   100 & 1000 &   100 000\cr

\+&&$\rho(x)$& Model&  .0013733& .0001491& .1148 & .1138 &
.01205 & 5.73$\s\s\times 10^{-6}$ &  1.30 
$\s\times 10^{- 12}$ & 1.14$\s\times 10^{-18}$&  7.94$\s\times
10^{-31}$\cr  
 
\+ &&&Weak&  1.813 & 1.137  & .002052  & .001396  & .0001995& 1.51$\s\times 10^{-6}$ &  1.51$\s\times
10^{-12}$ & 1.51$\s\times 10^{-18}$ &   1.19$\s\times
10^{-30}$\cr
\b }}

3. The gravitational mass is given by
$$
2m = \lim x\lambda(x) = 2.379,~~n = 6,...,100.
$$
  This is not the integral of a
mass density or an energy density, although there is an expression for this
number in terms of an integral
$$
2m = 2M(\infty)  = \int_0^\infty(\e^{-\nu}\rho-p)x^2dx + M(0). 
$$
The numbers for $n = 6$ are $2m = 2.379 = 4.303 +(-1.924 )$. 
The integral is not the integral of a density. 
  \b
  
  To give a  sense of the effect of larger values of $n$ we  
present the following Table 3, showing  the maxima of the function
$g^{00} =
\e^{-\nu}$, the functions  $\lambda$  and the density field $\rho$.

{\stephalf 
\b\ce{\bf Table 3 }

\settabs
 \+ :  ~~~~~~~~~~~~~~~~~~~~~ ~~~~~~~~~~~~~~~~~~~~ &6~~~~~~~~~~~~~~~&8~~~~~~~~~~
&10~~~~~~~~~~ &20~~~~~~~~~~~ &50~~~~~~~~~~ &100~~~~~~~~&1000\cr

 \+ &n:~~~~~~~~~~~~~~~~~ &6~~~& 10~~~ &20~~~ & 100~~ \cr
\+& Max($\s\rho$) &.1160 &   .2353&.5162&  2.6526&  
 \cr

\+&Max($\e^{-\nu}$)  & 2.630&  2.9036&  3.0317  
&3.040&  \cr  
 
\+& ~~~at $x$ = &3.086 &      3.2613&  3.3811&  
3.5125& \cr
 
\+&Max($ \s\lambda)$=&.6537& .8326&.9650& 1.0717& \cr
\+ &~~~at $x$ = & 4.432& 4.012&3.76815& 3.6055 & \cr
  \b}

\no The position of the density bump does not change much, but the
amplitude shows a strong rise. Since the total gravitational mass  remains
the same there must be concentration of mass close outside the
Schwarzschild radius. This is clearly illustrated by Fig.s 3a-c,
where one sees a dramatic narrowing of the bump at about 1.5 times the
Schwarzschild radius. 

\vskip -.5in
\def\picture #1 by #2 (#3){
  \vbox to #2{
    \hrule width #1 height 0pt depth 0pt
    \vfill
    \special{picture #3} 
    }
  }
\def\scaledpicture #1 by #2 (#3 scaled #4){{
  \dimen0=#1 \dimen1=#2
  \divide\dimen0 by 1000 \multiply\dimen0 by #4
  \divide\dimen1 by 1000 \multiply\dimen1 by #4
  \picture \dimen0 by \dimen1 (#3 scaled #4)}
  }

\parindent=1pc


\parindent=1pc

\vglue2cm 
\epsfxsize.8\hsize
\centerline{\epsfbox{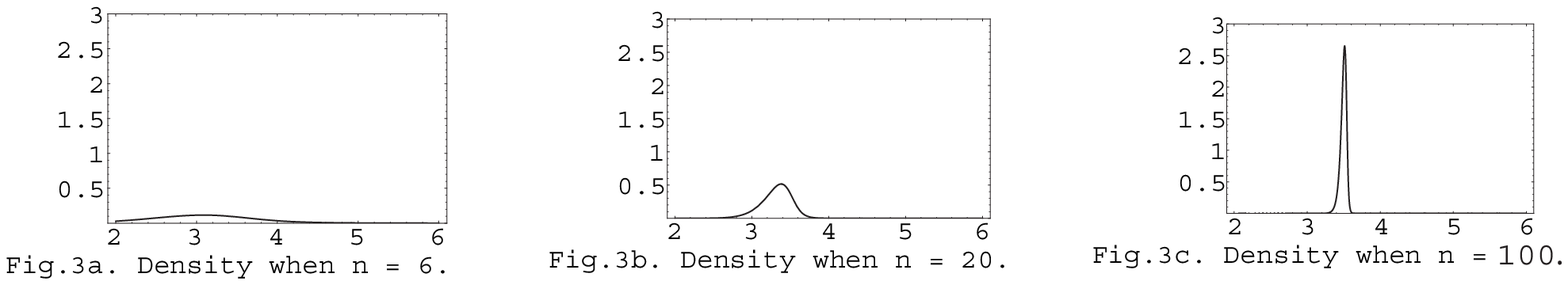}}
\no{\it Fig.3. The density peak narrows as $n$ is increased.}

\b 

4. Finally, Fig.s 4a-c illustrate  the gradual approach of the function
$g^{00} = \e^{-\nu}$ towards the Schwarzschild limit as $n$ grows. This
function, for large $n$, has a very sharp bend where it joins up with the
Schwarzschild metric.  Matter is strongly concentrated at this point
and the density is very low inside. (Fig.4d will be explained later.)

As is seen from these last figures, we have not been completely successful
in finding a limiting equation of state that reproduces the Schwarzschild
metric all the way down to the Schwarzschild radius. The association of
such a limit with  large polytropic index seems, however, to be
confirmed. A closer apprach  will be obtained
later.
 
We think that the results  support our contention that the
Schwarzschild metric should be interpreted in terms of a stiff equation
of state and a  concentration of matter at $ x\leq 2m$.
 
\b 
\vskip1.1in
\def\picture #1 by #2 (#3){
  \vbox to #2{
    \hrule width #1 height 0pt depth 0pt
    \vfill
    \special{picture #3} 
    }
  }
\def\scaledpicture #1 by #2 (#3 scaled #4){{
  \dimen0=#1 \dimen1=#2
  \divide\dimen0 by 1000 \multiply\dimen0 by #4
  \divide\dimen1 by 1000 \multiply\dimen1 by #4
  \picture \dimen0 by \dimen1 (#3 scaled #4)}
  }

\parindent=1pc

\vskip-3cm
\epsfxsize.8\hsize
\centerline{\epsfbox{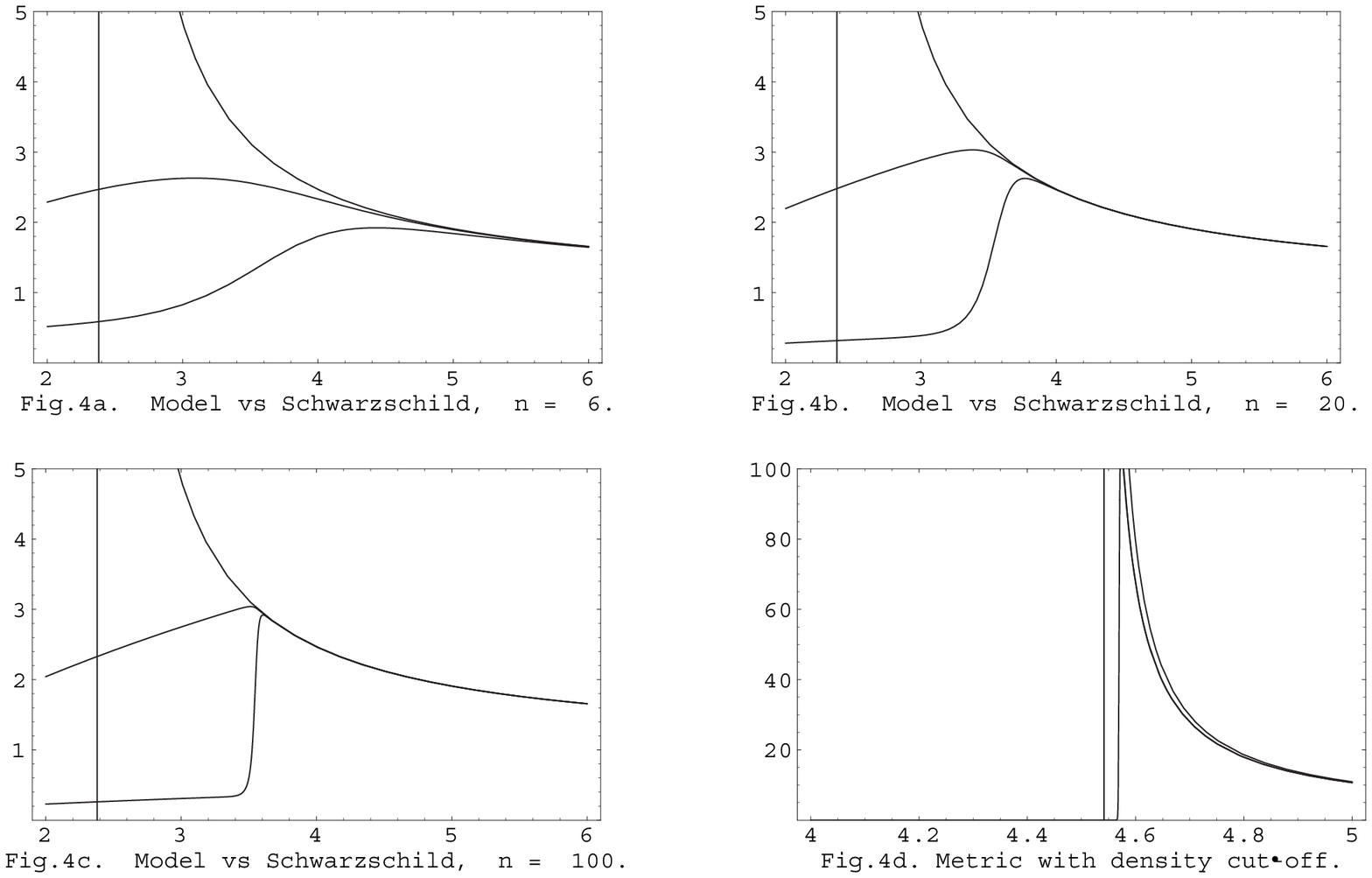}}
\vskip-1cm
\vskip1cm
\no{\it Fig.4. As $n$ is increased $g^{tt}$ and $g_{rr}$ cling closer to the
Schwarzschild metric. In the last figure the effect is enhanced by making the
radius finite.
}    

\vskip1cm

\ce{\bf Conclusions, infinite stars}

The main result of this investigation of infinite stars with finite mass is that
they seem to have a chance to exist.  The metric is smooth all the way to the
center. The low density found in the interior is interesting, although the
persistence of a small central region in which the pressure turns negative is
disturbing. The most optimistic explanation is
that it results from adopting a somewhat naive equation of state, with the same
polytropic index throughout. Perhaps the correct attitude to this problem was defined by
Chandrasekhar [C]: ``It is of course clear that if we agree to describe the
outer parts of a configuration by   singularity possessing solutions of the
differential equations involved, then we must assume that somewhere in the inner
regions the perfect gas laws break down." 

 \b\b

\no{\steptwo 6. Finite  stars }

Eddington adapted the polytropic model to stars with boundary. It is
useful to return for a moment to the weak field approximation.
\b
\ce{\bf Weak fields}

 For a finite matter distribution the restriction
$ n > 3$ is no longer relevant. Equation (4.3) was integrated numerically
for many values of $n$   by Emden  [Em], and
some of the results are quoted by Eddington [Ed].  
   As we pointed out in connection with 
Eq.s (3.1-3) the last of those equations, namely Eq.(3.3), 
$$
{1\over 2}(\e^{-\nu} - 1) = {dV\over d\rho}\eqno(6.1)  
$$
is not normally invoked in the traditional approach; instead the
`hydrostatic condition' 
$$
-(\e^{-\nu})'/2 =  p'/\rho,\eqno(6.2)
$$ 
is used, or the weak field approximation $\nu'/2 = p'/\rho$.. However, here
Eddington integrates this equation and  chooses the integration constant (which
is regarded as insignificant) so as to end up with (6.1) precisely. 
Consequently,   our equations reduce exactly to  those used by Eddington
  in the approximation  of weak fields.   This approximation was always taken
to be justified since the metric fields in  most stars (except
neutron stars and hypothetical super massive stars) are in fact weak.
See, however, Remark 1 at the end of Section 4.  

\b
\ce{\bf Boundary conditions }

In accord with the
principle that the gravitational field is determined by the mass that lies
closer to the origin, Eddington looked for solutions that are finite at the
origin and satisfy
$\nu'(0) = 0$. To apply the theory to a finite star,
Eddington 
considered the case
 
\no  $n < 5$, to take advantage of the fact that, in this
case, solutions of  Eq.(4.3),
$$
f'' + {2\over  r}f' + k^2 f^n = 0 
$$
that are finite and positive at the origin pass through zero at a finite
value of the radius. Since  the pressure turns negative, this point must lie
outside, or at the boundary, of the region where the theory is applicable.
Eddington supposed that the first zero of the pressure indicates the boundary
of the star. This implies that the gravitational potential also vanishes at
the surface, and because  Eddington's choice ot the   constant in
his integration of (6.1) brings him into agreement with the wave equation
Eq.(3.3), we can identify his gravitational potential with $-\nu/2$. However,
as  already pointed out  in the preceding section, Eddington cannot 
fix a solution by matching for $g_{00} = \e^\nu$  to a
Schwarzschild metric on the outside, which is why he must impose  boundary
conditions at the origin, thus having to  estimate   the conditions that are
likely to prevail at the center of the star. The `mass' is defined as the
integral of the density. In the nonrelativistic case this definition of the mass
agrees with our definition in terms of the asymptotic metric.

An essential feature of our model is that the integration of the
hydrostatic equation is determined without ambiguity and matching up to an
outside metric is no longer trivial. Outside the star the metric is  
that of   Schwarzschild, and in particular
$\nu +
\lambda = 0$. In the weak field approximation $\lambda$ is eliminated
by means of the relation $\lambda = r\nu'$, so we must have
$$
r\nu'(r) + \nu(r) = 0, ~r \geq R.
$$  
This implies the boundary condition
$$
Rf'(R) + f(R) = 0,\eqno(6.3)
$$ 
which is incompatible with $f(R) = 0$.
\ve
The equation of state that is
being used in the interior is not expected to apply to matter near the surface. 
The idea that the pressure drop  continuously to zero at the boundary
is reasonable,  but we are not sure that the same can be said  of
the density. 
Oppenheimer and Volkoff, who also fix the boundary of the star at the point where
the pressure drops to zero, seem to agree with this, for they say that there are
many equations of state that allow the pressure to go to zero  
at finite density. The one they use does not have that property [VO].  In
order to compensate to some degree for the shortcomings of the polytropic
equation of state    we   take a positive value for
$f(R)$. Since the boundary value of $f'(R)$ is fixed in terms of
$f(R)$, we shall get a 1-parameter family of solutions.

This applies to the weak field approximation as well as to the model.
In the case that $n$ = 1, $\gamma = 2$ Eddington's solution is
$$
\nu(x) = - { c\over x}\sin{x\over 2},~~ X =  2\pi,~~ c = \int r^2\rho(x)dx),
$$
where $x = X$ is the boundary, corresponding to $r = R$.
Our solution, in the weak field approximation (valid in this case) and with the
boundary conditions  imposed by the model, is\break 
$$
\nu(x) = -{2m\over x}\sin{x\over 2},~~ X =  \pi,   
$$
where $m$ is the gravitational mass.

This result is significant. In this particular case, when $n = 1$  and in the
weak field approximation, the boundary condition (6.3), that comes from the
matching of the interior metric to the exterior Schwarzschild metric,   
unexpectedly  selects the only solution that is regular at
the origin, thus vindicating, in this case at least, some of Eddington's
precepts.

It was found, empirically, that this last observation applies to the
exact solutions of the full model as well. Namely, if the parameters $R$ and
$\nu(R)$ are not chosen with special care, then the solution has a very singular
behaviour at the origin. This suggests that Eddington is right in
choosing regular boundary conditions at the origin; this also guarantees that
pressure and density both remain positive throughout. The least that can be
said is that the stars that satisfy this criterion form a distinguished
family, characterized by a critical relationship between radius and
gravitational mass. We shall see that this implied relationship between radius
and mass is not one-to-one; for fixed values of $n$ and $R$ there are from 1
to 3 possible values of the mass. 

\b
\ve

\ce{\bf Numerical results for finite stars}

Fig.s 5a-d show examples of the contrast between Eddingtons star, where density
and pressure both tend continuously to zero at the boundary, and the solution
obtained from the model, matched to Schwarzschild at the boundary, where density
and pressure both drop abruptly to zero. These are the boundary conditions that
we impose from now on. 

\vskip1.5in
\def\picture #1 by #2 (#3){
  \vbox to #2{
    \hrule width #1 height 0pt depth 0pt
    \vfill
    \special{picture #3} 
    }
  }
\def\scaledpicture #1 by #2 (#3 scaled #4){{
  \dimen0=#1 \dimen1=#2
  \divide\dimen0 by 1000 \multiply\dimen0 by #4
  \divide\dimen1 by 1000 \multiply\dimen1 by #4
  \picture \dimen0 by \dimen1 (#3 scaled #4)}
  }

\parindent=1pc

\vskip-3cm
\epsfxsize.8\hsize
\centerline{\epsfbox{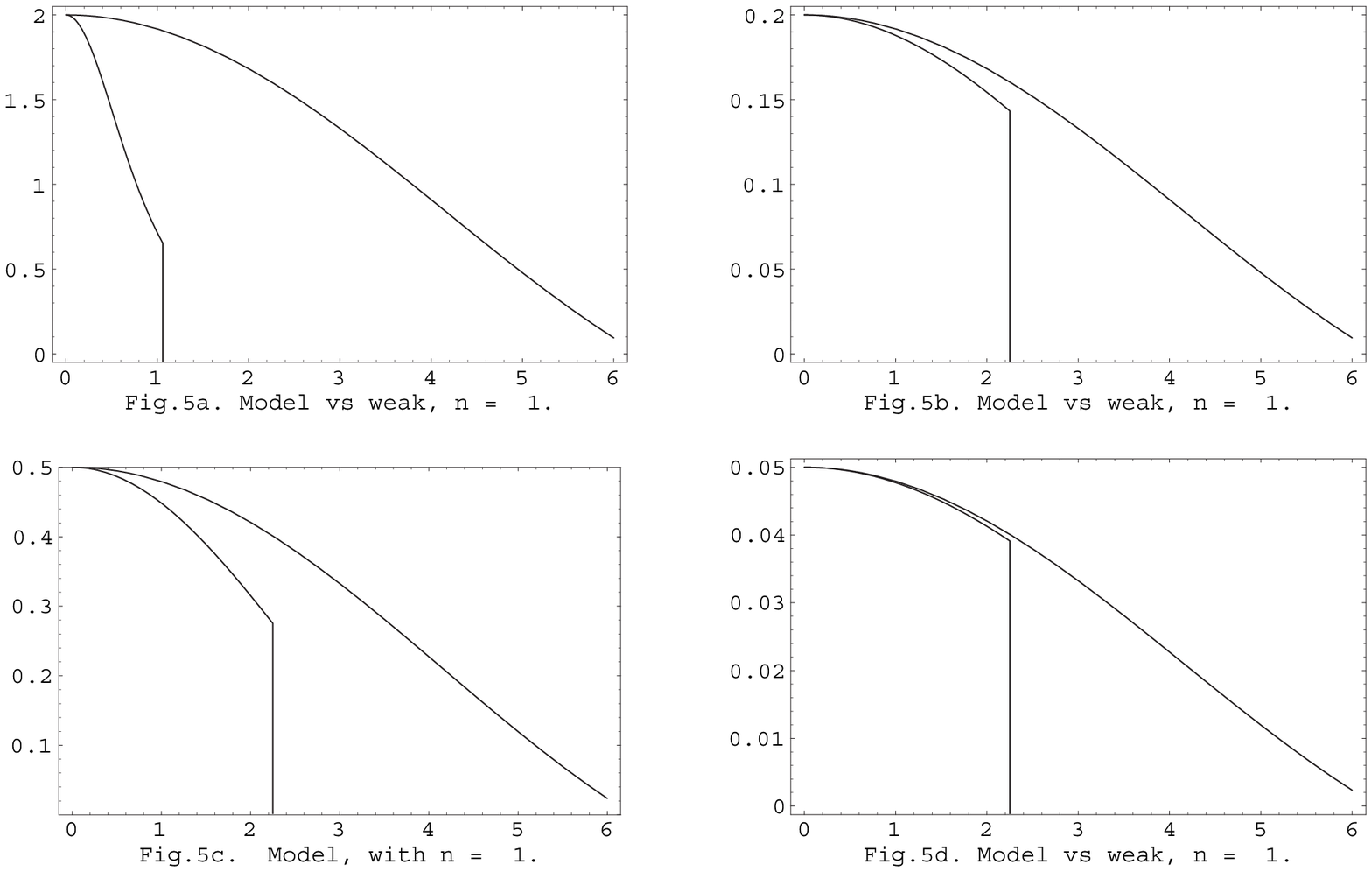}}
\vskip-1cm
\vskip1cm
\no{\it Fig.5. The pressure, for different central pressures. Comparison between
Eddington's star, with boundary at
$x = 6$ and the result of the model, which puts the boundary at the end of the
short curve, where the metric is matched to the external Schwarzschild metric.
}

\vskip1cm

 There is a critical, empirial relationship between the radius
$R$ and the
 gravitational mass, both being determined by the conditions of matching the
Schwarzschild metric at $r = R$,
$$
 \nu(R) = -\lambda(R) = \ln  (1-{2m\over R}),
$$ 
that leads to a metric that is regular at the origin. It is found  that most solutions
(of the exact equations of the model, as in the weak field approximation) blow
up at the origin, and that the exceptional case when
$\nu$ remains bounded can be localized to a very high degree of accuracy by the
condition that $\nu'(0)$ vanish. Thus to find these particular solutions we  
adopt  Eddington's boundary condition $\nu'(0) = 0$. Starting
the integration at the origin (actually at $x = 10^{-10}$),  varying the
value of $\nu(0)$ from .001 up to the maximum value  that the computer  program
allows, we integrate up to the first zero of $\nu + \lambda$, where the
star ends, to obtain a parametric representation of the critical relation between
$R$ and $2m$. The results are presented  in Fig.6a-d as a log-log plot  and
juxtaposed in Fig.7.

 \vskip1.1in
\def\picture #1 by #2 (#3){
  \vbox to #2{
    \hrule width #1 height 0pt depth 0pt
    \vfill
    \special{picture #3} 
    }
  }
\def\scaledpicture #1 by #2 (#3 scaled #4){{
  \dimen0=#1 \dimen1=#2
  \divide\dimen0 by 1000 \multiply\dimen0 by #4
  \divide\dimen1 by 1000 \multiply\dimen1 by #4
  \picture \dimen0 by \dimen1 (#3 scaled #4)}
  }

\parindent=1pc

\vskip-3cm
\epsfxsize.8\hsize
\centerline{\epsfbox{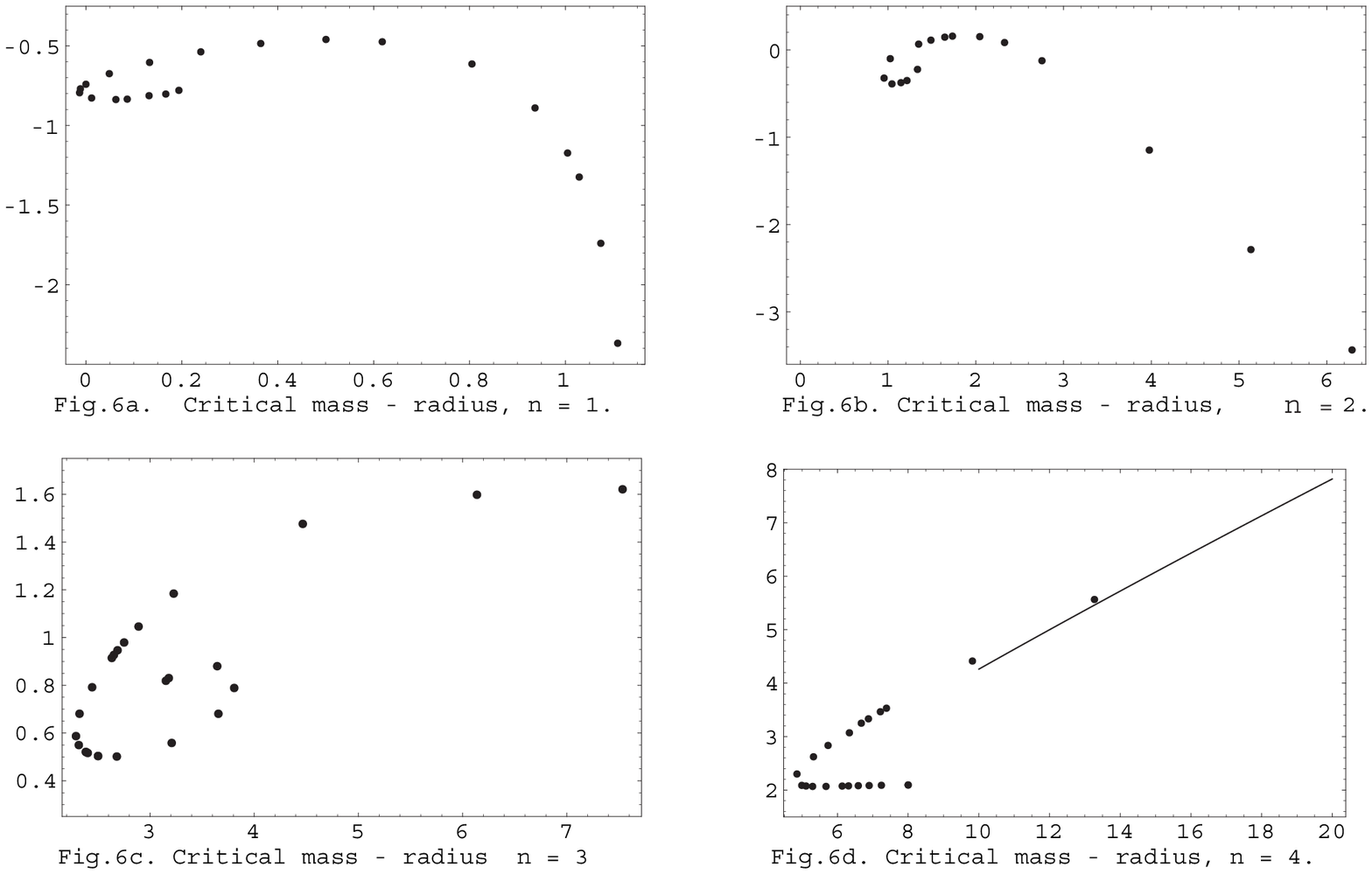}}
\vskip-1cm
\vskip1cm
\no{\it Fig.6a-d. The relation between mass and radius in a log-log plot. The abscissa
is the radius and the ordinate is $2m$.
}    
\vskip.1in
\def\picture #1 by #2 (#3){
  \vbox to #2{
    \hrule width #1 height 0pt depth 0pt
    \vfill
    \special{picture #3} 
    }
  }
\def\scaledpicture #1 by #2 (#3 scaled #4){{
  \dimen0=#1 \dimen1=#2
  \divide\dimen0 by 1000 \multiply\dimen0 by #4
  \divide\dimen1 by 1000 \multiply\dimen1 by #4
  \picture \dimen0 by \dimen1 (#3 scaled #4)}
  }

\parindent=1pc

\epsfxsize.8\hsize
\centerline{\epsfbox{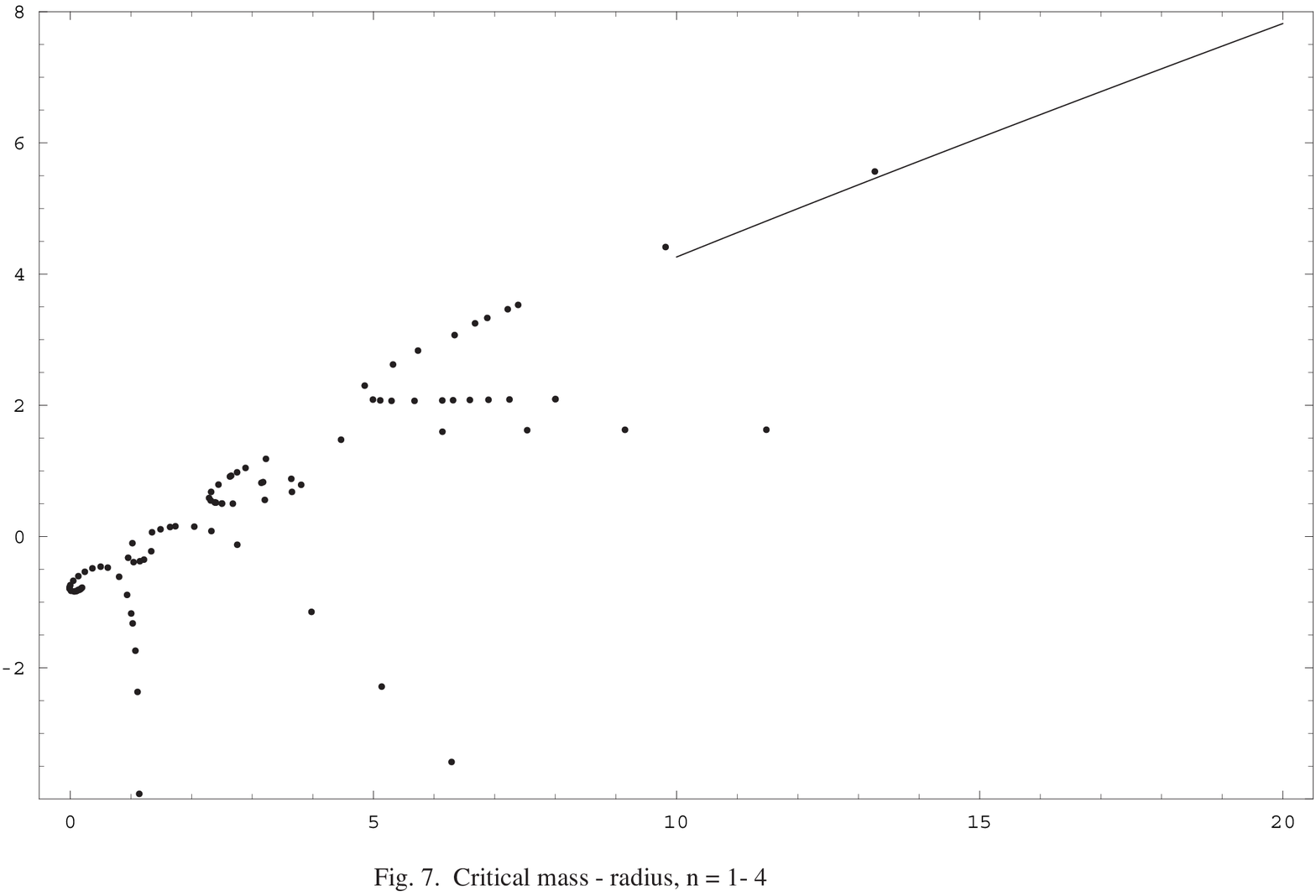}}
\vskip-1cm
\vskip1cm
\no{\it Fig.7. A composite of Fig.6a-d.     }    

\vskip1cm
For $n = 4$, in  the upper part,$  -\nu(0)\geq .4385$,  the
relation of
$X$ to
$m$ is well represented by  ln$(2m) = .567$ (ln$X)^{.876}$. In the figures  
it is represented by this curve, to distinguish this branch from
the other, in the region where they nearly coincide.

The strange shape of the curves gives rise to several different configurations
with the same gravitational mass but  different size.  Table 4 gives an example
for $n = 3$.

\b{\stephalf
\ce{\bf Table 4.}
\settabs \+~~~ x~~~~~~~~~~~~~~~~~~~~~~~~~~~~~~~~~~~~~~~~~~~~~~~& 10~~~~ ~~ 
~~~~& 20~~~~~~~   ~~~& 100~~~~~~~~~& 1000~~~~~~~& 10000~~~~~~&
100000~~~~~&~~~~~~~~~~~~~~~&~~~~~~~~~~~~~~~&\cr
\+& $\s 2m$& 2.49&2.41&2.21\cr

\+&$\s X$& 38.3&24.0&9.85\cr

\+&$\s -\nu(0)$&1.5&2.0&.7\cr}
\b
\no There are stars intermediary between the first two, both of which are unstable, suggesting aging with
small mass loss, but none between the last two, which indicates a cataclysmic
event. The star with the smallest radius is stable. The number $-\nu(0)$ is a measure of the temperature at the center.

\b
\no{\bf Neutron stars and  Oppenheimer-Volkoff theory }

In their ground breaking paper [OV] on neutron cores, Oppenheimer and Volkoff
used a very stiff equation of state. In order to to compare the predictions of
their theory with our model we shall have to do so with the same equation of
state and choose for this purpose the polytrope with $ n = 1/10$. Numerical
results are presented in Table 5.
\b

{\stephalf
\ce {\bf Table 5. $n = 1/10$.}
 
\settabs \+~~~~~~~~~~~~~~~~~~~~~~~ x~~~~~~~& 10~~~~ ~~  ~~~~& 20~~~~~~~   ~~~&
100~~~~~~~~~& 1000~~~~~~~& 10000~~~~~~&
100000~~~~~&~~~~~~~~~~~~~~~&~~~~~~~~~~~~~~~&\cr
 
\+& &Model &&& Oppenheimer-Volkoff [OV]\cr

\+ &-$\s\nu(0)$& .001&.01&.097&.001&.01&.097\cr 

\+&$\s X$&.07439&.20855&.5512& \cr

\+& -$\s \nu(X)$ & .0006637&.00613&.062&0.00067 &.00656&.06329\cr

\+&$\s 2m$& .00004936&.0013746&.0331& .000049&.00137&.03376\cr

\+& $\s\rho(0)$&.3679&.4634&.5842&.3679&.4634&.5842\cr

\+& $\s\rho(X)$&.3532&.4445&.5576&.3532&.4448&.5588 \cr

\b}
 
The most significant parameter is $2m/X$, the potential at the surface relative
to the maximum possible that is reached when the surface of the star coincides
with the Schwarzschild horizon. The highest value found  in our model for this
equation of state ($n = 1/10$) is only .06. This explains the near coincidence of
the results; the gravitational fields are weak.
 
The big contrast between the two theories is that the model imposes stronger
conditions for matching up with the external Schwarzschild metric, to wit, the
condition $\nu(X) + \lambda(X) = 0$. According to Oppenheimer and Volkoff the
boundary of the star is not at the point $x =X$, and indeed $X$ is not defined
in their theory, but at the first zero of the density. (Our computer program did
not allow us to carry the integration that far.) The values $\nu(X),...,\rho(X)$
on the right side were read at the corresponding values of $X$ computed from the
model. 

The question of the greatest possible mass for a neutron star is too
complex to be raised here; see [H]. We  venture only to propose that
the upper limit on the parameter
$2m/X$ is about .06 for small $n$,   but   values up to 2/3  are possible with
polytropic index $n = 1$. 
\b\b

\no{\steptwo Acknowledgements}

I thank Jim Hartle for very useful, critical  comments. I am grateful to Zwi Bern, Michael Gutperle, 
  Prakashan Korambath and Gilda Reyes for assistance.

 \ve

\no{\steptwo References.}
\settabs 
\+&~~~~~~~~~~&\cr

 \+ &[C]&   Chandrasekhar, S., Stellar Configurations with Degenerate Cores,
\cr

\+ & &   Monthly Notices R.A.s., {\bf 95}, 226-260 (1935).\cr

\+&[D]& Dirac, P.A.M., Particles of finite size in the gravitational field,\cr
 
\+&&Proc. Roy. Soc. London {\bf A270},  354-356 (1962),
\cr

\+&[Ed]& Eddington, A.S., {\it The Internal Constitution of Stars}, Dover,
N.Y. 1959.    \cr

\+&[Em]& Emden, R., {\it Gaskugeln}, Teubner 1907. \cr

\+&[FW]& Fetter, A.L. and Walecka, J.D., {\it Theoretical Mechanics of Particles
and Continua}, \cr

\+&[Fo]& Fowler, R.H., On dense matter, Ap.J. {\bf 144},
180 (1935).,\cr

\+&&McGraw-Hill 1980. QA 808.2F47.
\cr

\+ &[Fr] & Fr\o nsdal, C., Growth of a Black Hole, Geom.  Phys., to
appear,  gr-qc/0508048.\cr

\+ &[H] &Hartle, J.B., Bounds on   mass and moment of inertia of non-rotating
neutron stars,\cr 

\+&&  Physics Reports,{\bf 46}, 201-247 (1978).\cr

\+&[KW] &Kippenhahn, R. and Weigert,A., {\it Stellar Structure and Evolution}, Springer-Verlag1990.\cr

\+&[L]&Lang, K.R., {\it Astrophysical Formulae}, Springer 1999.   \cr

\+& [M]& Milne, E.A., Thermodynamics of the stars, in {\it Handbuch der
Astrophysik} {\bf 3}, 80 (1930). \cr

\+&[MTW]&Misner, C.W., Thorne, K.S. and Wheeler, J.A., {\it Gravitation},\cr

\+&&Freeman, San Francisco 1970.\cr

\+&[OV]& Oppenheimer, J.R. and Volkoff, G.M., On Massive Neutron Cores,\cr 

\+&&Phys.Rev.
{\bf 55}, 374-381 (1939).\cr

\+&[S]& Shutz, B.F.Jr., Perfect fluids in General Relativity: Velocity
potentials and a \cr

\+&&variational principle, Phys.Rev.D {\bf 2}, 2762-2771
(1970).\cr

\+&[SW]& Seliger, R.L. and Whitman, G.B., Proc.Roy.Soc. {\bf A305}, 1 (1968).\cr

\+&[Ta]& Taub, A.H., General relativistic variational principle for
perfect fluids,\cr

\+&& Phys.Rev. {\bf 94}, 1468 (1954).\cr

\+ &[To]&  Tolman, R.C., {\it Relativity, Thermodynamics and Cosmology},
Clarendon, Oxford 1934.\cr

 \+& [W]& Weinberg, S., {\it Gravitation and Cosmology}, Wiley, 1972.\cr
\b  
\end
\no{\steptwo Figure Captions}

Fig 1. On the left, $\e^{-\nu}$ on top and $\e^\lambda$ below. On the right,
the density. Parameters $n = 6$, $2m$ = .2,~ 2.5 or 10.2  as indicated.

Fig.2. Comparison between the model, the ``weak field approximation" and the
Schwarzschild metric in 4 cases. The hyperbola is the Schwarzschild metric,
$g^{tt} = g_{rr}$. The lower curve is $g^{00}$ in the model. The upper curve,
with the sharp peak inside the Schwarzschild radius, is the same function in the
``weak field approximation".

Fig.3. The density peak narrows as $n$ is increased.

Fig.4. As $n$ is increased $g^{tt}$ and $g_{rr}$ cling closer to the
Schwarzschild metric. In the last figure the effect is enhanced by making the
radius finite.

Fig.5. The pressure, for different central pressures. Comparison between
Eddington's star, with boundary at
$x = 6$ and the result of the model, which puts the boundary at the end of the
short curve, where the metric is matched to the external Schwarzschild metric.

Fig.6a-d. The relation between mass and radius in a log-log plot. The abscissa
is the radius and the ordinate is $2m$.
 
Fig.7. A composite of Fig.6a-d.

\end